# Localization of Seizure Onset Zone based on Spatio-Temporal Independent Component Analysis on fMRI

Seyyed Mostafa Sadjadi [1,2], Elias Ebrahimzadeh [1,2,*], Alireza Fallahi [1,3],
Jafar Mehvari Habibabadi [4], Mohammad-Reza Nazem-Zadeh [5,6,7], Hamid Soltanian-Zadeh [1,2,8]

[1] Control and Intelligent Processing Center of Excellence (CIPCE), School of Electrical and Computer Engineering, University of Tehran, Tehran, Iran.

[2] Neuroimage Signal and Image Analysis Group, School of Cognitive Sciences, Institute for Research in Fundamental Sciences (IPM), Tehran, Iran.

[3] Biomedical Engineering Department, Hamedan University of Technology, Hamedan, Iran.

[4] Isfahan Neuroscience Research Center, Isfahan University of Medical Sciences, Isfahan, Iran.

[5] Research Center for Molecular and Cellular Imaging, Advanced Medical Technologies and Equipment Institute (AMTEI), Tehran University of Medical Sciences, Tehran, Iran.

[6] Medical Physics and Biomedical Engineering Department, Tehran University of Medical Sciences, Tehran, Iran.

[7] Department of Neuroscience, Monash University, Melbourne, VIC, Australia.

[8] Medical Image Analysis Laboratory, Departments of Radiology and Research Administration, Henry Ford Health System, Detroit, MI, United States.

**\* Corresponding Author:**
**Elias Ebrahimzadeh,**
1. Control and Intelligent Processing Center of Excellence (CIPCE),
School of Electrical and Computer Engineering,
College of Engineering, University of Tehran,
North Kargar Ave., Tehran, Iran.
P.O. Box: 14395-515
Email: e_ebrahimzadeh@ut.ac.ir
2. Neuroimage Signal and Image Analysis Group,
School of Cognitive Sciences,
Institute for Research in Fundamental Sciences (IPM),
Niavaran Ave., Tehran, Iran.
P.O. Box: 19395-5746
Email: elias.ebrahimzadeh@ipm.ir
ORCID: 0000-0001-8682-936X

# Abstract

Localization of the seizure onset zone (SOZ) as a step of presurgical planning leads to higher efficiency in surgical and stimulation treatments. However, the clinical localization including structural, ictal, and invasive data acquisition and assessment is a difficult and long procedure with increasing challenges in patients with complex epileptic foci. The interictal methods are proposed to assist in presurgical planning with simpler data acquisition and higher speed. In this study, spatio-temporal component classification is presented for the localization of epileptic foci using resting-state functional magnetic resonance imaging (rs-fMRI) data. This method is based on spatio-temporal independent component analysis (ST-ICA) on rs-fMRI with a component-sorting procedure upon dominant power frequency, biophysical constraints, spatial lateralization, local connectivity, temporal energy, and functional non-Gaussianity. This method aimed to utilize the rs-fMRI potential to reach a high spatial accuracy in localizing epileptic foci from interictal data while retaining the reliability of results for clinical usage.

Thirteen patients with temporal lobe epilepsy (TLE) who underwent surgical resection and had seizure-free surgical outcomes after a 12-month follow-up were included in this study. All patients had pre-surgical structural MRI and rs-fMRI while post-surgical MRI images were available for ten. Based on the relationship between the localized foci and resection, the results were classified into three groups "fully concordant", "partially concordant", and "discordant". These groups had the resulting cluster aligned with, in the same lobe with, and outside the lobe of the resection area, respectively. This method showed promising results highlighting valuable features as SOZ functional biomarkers. Contrary to most methods which depend on simultaneous EEG information, the occurrence of epileptic spikes, and the depth of the epileptic foci, the presented method is entirely based on fMRI data making it independent from such information and considerably easier in terms of data acquisition, artifact removal, and implement.

# 1. Introduction

Epilepsy is one of the most widespread neurological disorders (Jobst and Cascino, 2015) causing spontaneous seizures as a sort of excessive, abnormal, or synchronous neuronal activities (Fisher et al., 2005). The seizures can originate from one or several specific zones or can generalize over the brain tissue. Using anti-seizure drugs, epilepsy can be controlled, but can also be drug-resistant in about twenty to thirty percent of the cases (Schuele and Lüders, 2008). For refractory epilepsy, surgical resection is among the well-established approaches to control seizures (Engel, 2018). However, presurgical planning should include a localization step for epileptic foci so the surgeon can use it as a perspective for opening the skull and the surgical progress leads to proceed more efficiently. The removal of epileptogenic brain tissue may lead to cognitive deficits particularly when the resection areas are associated with working memory, attention, or executive functions. Targeted rehabilitation helps in regaining lost functions or compensating for them through neural plasticity (Asgarinejad et al., 2023). Cognitive therapies and neurostimulation techniques have shown promise in enhancing recovery and improving patients' quality of life post-surgery, allowing better integration into daily activities and social environments (Asgarinejad et al., 2024).

Various methods including ictal, interictal, invasive, and non-invasive data recording are developed to localize the generation area, spatial extent, and propagation pathways of epileptic activity with a constant challenge of balance between accuracy and simplicity of data acquisition and implementation (Baumgartner et al., 2019). The current gold standard for localizing epileptic foci is from ictal (seizure) onset, which requires the occurrence and recording of multiple typical seizures of a patient (Vlachos et al., 2017), and the frequency of seizure occurrence is relatively low compared with interictal epileptiform discharges (IEDs). Invasive approaches like intracranial electroencephalography (iEEG) recording provide high yet local spatial resolution and hence have attracted attention to define the epileptogenic zone (EZ) and localize the seizure focus in later steps of presurgical planning (Rosenow and Lüders, 2001). Despite the time-consuming process of the clinical protocol, the success rate of the resection surgery varies depending on several factors. Post-surgical follow-ups can be from a few months to more than five years, and the surgical outcomes are divided into four classes according to the Engel criteria (Engel J et al., 1993) from Engel I to Engel IV corresponding to seizure freedom (about 50-70% of cases), with warning signs or minor seizures for less than 3 days per year (about 10-30% of cases), greater than 80% reduction in seizure frequency or worthwhile improvement in the seizure-related disability (about 10-30% of cases), and finally less than 80% reduction in seizure frequency or no worthwhile improvement (less than 10% of cases), respectively (Engel, 2018).

Although the localization of the epileptic foci from seizure-free (interictal) periods remains a challenging problem, especially in the absence of IEDs, the non-invasive methods for that have become under greater attention over the past few years, mostly the ones based on EEG-correlated fMRI (EEG-fMRI) (Sadjadi et al., 2021a). This combination benefits from the high temporal resolution of the EEG signal and the high spatial resolution of blood oxygen level-dependent (BOLD) fMRI at the same time. Khoo et al. (Khoo et al., 2018) showed that the IED adjacent to a maximum BOLD response, which often corresponds to the seizure onset zone, is more likely to precede IEDs in remote locations during a widespread intracranial discharge. Therefore, simultaneous EEG-fMRI is a unique non-invasive method to reveal the origin of IEDs. It is noticeable that the localization of epileptic foci using these approaches is a key step of presurgical planning which helps surgeons have a perspective on the approximate region of interest for opening the skull, although the resection process will be done with a precise investigation using intracranial electrodes during surgery.

In the localization of epileptic foci, combining two neuroimaging modalities has generated more accurate results than a single modality (Ebrahimzadeh et al., 2019b, 2019a). Lei et al. (Lei et al., 2011) showed that localizing based on EEG data alone even using independent component analysis (ICA) leads to poor results. However, localization of epileptic foci based on fMRI data alone is still under attention. Several studies argue that the relationships between BOLD and local field potential (LFP) are not always linear and may change depending on various factors. It has been shown that functional connectivities measured by BOLD and EEG signals have relatively weak correlations (Bettus et al., 2011). Not a lot of studies have investigated the neural basis of such spontaneous fluctuations in fMRI signals (He et al., 2008; Shmuel and Leopold, 2008). These two modalities measure different phenomena related to epilepsy possibly occurring at different time scales. We might see specific electrophysiological features of epileptic networks in stereotactic EEG (SEEG), while rs-fMRI normally reflects the functionality of such networks (Bettus et al., 2011).

The influence of EEG spikes on BOLD signals is not quite clear (Bettus et al., 2011). Simultaneous EEG-fMRI recordings have revealed that spikes can either lead to increased, decreased, or unchanged BOLD signals (Bénar et al., 2006). The hypothesis of neurovascular decoupling has also been questioned to explain this complexity (Lemieux et al., 2008). EEG-fMRI is generally limited by the detection of frequent spikes on scalp EEG and the underestimation of those not expressing properly on surface EEG may be due to a source in deep brain structures (Bettus et al., 2011; Ebrahimzadeh et al., 2021, 2022). This can cause a false BOLD signal baseline. IEDs may themselves be at least in part responsible for the discrepancy between EEG and BOLD coupling, but already found negative correlations between both these signals in regions spared by epileptiform abnormalities claim that spikes are not solely responsible (Bettus et al., 2011).

There are useful tools to analyze and study abnormal activities with specific sources in brain volume. The idea of exploring fMRI data for candidate sources related to epilepsy was discussed by Zhang et al. (Zhang et al., 2015) using a set of selection criteria on the components obtained from spatial ICA (sICA). Temporal clustering analysis (TCA) (Morgan et al., 2004) and temporal ICA (tICA) are also useful to find and confirm the epileptic activity among epileptogenic sources (Chen et al., 2006). Moreover, functional connectivity analysis is a promising way to look for the functionally integrated relationships throughout a cluster and between spatially separated brain regions, which can be helpful to define local network topological features, localize the significantly connected areas to the SOZ, and the propagation pathways of epileptic activity (Su et al., 2019). In Table 1, pivotal studies that present interictal methods for epileptic foci localization using EEG, fMRI, and EEG-fMRI data and include seizure-free surgical outcomes for evaluation are reported chronologically. It is noteworthy that the mentioned methods might have been used and evaluated in other studies after their presentation as well on different groups of patients.

According to Table 1, localization methods using interictal data have been being tested for several years. From a major investigation (Sadjadi et al., 2021b), only 23% of the presented methods in the literature have been evaluated with seizure-free surgical outcomes. The evaluation criterion in these studies is majorly the visual matching of the detected seizure focus with the true SOZ. The average accuracy of the presented studies in Table 1 is 61%, which is equivalent to a total of 152 concordant results out of 250 subjects who underwent surgery and were free from epileptic attacks.

In this study, a novel method is presented to localize the epileptic foci using fMRI data, which is easily implemented and avoids the previously mentioned disagreements. Our dataset included the fMRI data of thirteen patients with temporal lobe epilepsy who were candidates for resection surgery. Ten patients had resection surgery with seizure-free outcomes after a follow-up of more than 12 months. The method presented in this paper is noninvasive in nature with no requirement of simultaneously recorded EEG signals. This mainly simplifies data acquisition, artifact removal, and implementation, in addition to not being involved with the mentioned questions about the basis of EEG-fMRI studies by being independent of the spike detection process and the foci depth.

## 2. Materials and Methods

The presented method in this study is based on ST-ICA on rs-fMRI data with a component-sorting procedure upon dominant power frequency, biophysical constraints, spatial lateralization, local connectivity, temporal energy, and functional non-Gaussianity.

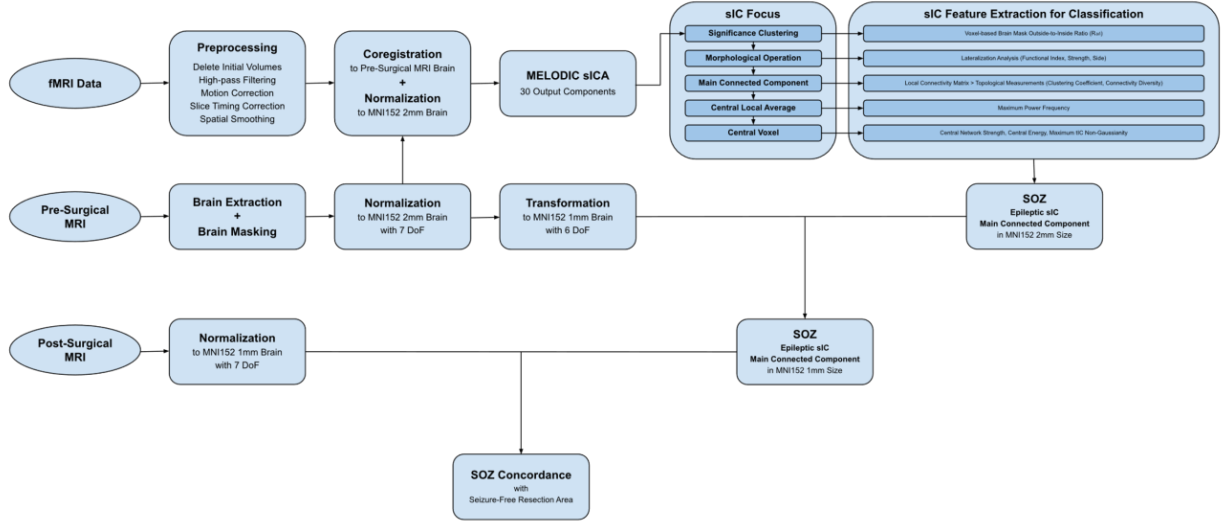

Figure 1 shows the diagram of the presented method in detail including each step of the whole analysis algorithm from the raw data until the SOZ localization and concordance assessment.

### 2.1. Patients and Data

This research was reviewed and approved by the Research Ethics Board (Institutional Review Board, IRB) of the Tehran University of Medical Sciences. Patients with severe cognitive impairment or other neurological diseases were excluded from the study. The patients with severe cognitive impairments at a level of preventing participation in the full structural, diffusion, or functional MRI study were also excluded. All volunteers signed informed consent to participate in the study. Thirteen unilateral TLE subjects who had undergone resection surgery and had seizure-free (Engel I) outcomes after a 12-month follow-up were recruited. The dataset includes pre-surgical MRI and fMRI data for all patients along with post-surgical MRI images for ten. All subjects were scanned using a 3-Tesla Siemens Magnetom Prisma MRI. Anatomical images were acquired for clinical diagnosis including transverse T1-weighted images with TR = 1840 ms, TE = 3.47 ms, flip angle = 8°, matrix = 256 × 256, and slice thickness = 1.0 mm. The rs-fMRI was acquired in transverse planes covering the whole brain with 330 measurements. The parameters of the echo planar imaging sequence included TR = 3000 ms, TE = 30 ms, flip angle = 90°, matrix =

640 × 640, slice thickness = 2.4 mm. For each subject, the duration of each fMRI measurement was approximately 16.5 minutes. All subjects were asked to keep their eyes closed during the resting state fMRI scanning process. Emphasis was placed on their not sleeping and confirmation was secured after the scan that they had remained awake throughout. The clinical and electrophysiological characteristics of the patients are available in Table 2.

## 2.2. Preprocessing

The fMRI data were preprocessed and analyzed using FSL (FMRIB Software Library, https://fsl.fmrib.ox.ac.uk/fsl) and Python. The first ten volumes were discarded to ensure steady-state magnetization. A high-pass temporal filter with a cutoff of 100 seconds was applied to fMRI data to eliminate the low-frequency drifts known as detrending. Head motion correction was performed via a 6-parameter rigid-body transformation based on the MCFLIRT algorithm. Temporal autocorrelations were corrected with an autoregressive model of order one. The data were spatially smoothed using a Gaussian filter with 6-mm full width at half-maximum (FWHM) for an increase in signal-to-noise ratio (SNR).

Functional images were registered to the pre-surgical structural MRI scan and then were normalized to the MNI152 brain template with a 2 mm voxel size using 7 degrees of freedom (DoF). For later concordance assessment, the full-skull post-surgical structural MRI scan was normalized to the non-brain-extracted MNI152 template with 1mm voxel size using 7 degrees of freedom (DoF) and mutual-information cost function to retain the shape of the resected brain region. Moreover, the transformation from normalized pre-surgical structural MRI scan to the MNI152 brain template with 1mm voxel size using 6 DoF was also calculated so the resulting SOZ can be transformed to high-resolution space for concordance assessment with normalized post-surgical structural MRI scan.

sICA was applied to fMRI data to explore and classify the candidate components associated with epilepsy (Sadjadi et al., 2022). Thirty components with variance-normalized timecourses were considered proper (Zhang et al., 2015) to first computed for each patient, and the voxel intensities of each sIC map were converted to z-scores representing the spatial distribution of each component. Five levels of clustering were considered for feature extraction of each sIC. Significance clustering required every cluster with more than ten contiguous voxels having Z-score > 3.1; morphological operated clusters were made after opening and closing with a disk structure of 2 voxels; the main connected component was the connected cluster including the sIC center with maximum Z-score voxel; the central local average was the averaged fMRI data in 3-voxel neighborhood of the sIC center; and the central voxel was the fMRI data of the sIC center.

## 2.3. Feature Extraction

For feature extraction, a set of potential functional biomarkers for epileptic focal activity were considered. This set is based on frequency features, biophysical constraints, spatial lateralization, local connectivity, temporal energy, and functional non-Gaussianity. These features were extracted as the indicators for the classification of epileptic sIC and localization of SOZ cluster as follows:

### 2.3.1. Maximum Power Frequency

Based on the expected temporal structure in each component for being neurophysiologically meaningful and normal (Cordes et al., 2001), the dominant frequency of each sIC's central local average was considered in our feature set. The averaged fMRI data was calculated in a 3-voxel neighborhood of the maximum Z-score voxel as the sIC center and a periodogram was used to find the dominant frequency. It is noteworthy that for not-preprocessed fMRI data, a dominant frequency higher than 0.1 Hz is considered a reflection of aliasing of respiration and cardiac artifacts, while lower than 0.01 Hz is considered scanner susceptibility artifacts. Therefore, the components with dominant power outside of the frequency range of 0.01–0.1 Hz should be excluded from further analyses.

### 2.3.2. Voxel-based Outside-to-Inside Ratio

According to the expectation of neurological activities being generated by neurons residing within the grey matter of the cortex, those components confounded with external sources of artifacts should have been excluded. To this aim, fMRI voxels outside of the brain, which are caused by noise, were retained before performing sICA. Having been included in the noisy components, they were used to identify and reject those noisy activities within the brain that have a statistical correlation with them. The index of $R_{o/i}$ was calculated as follows to discriminate the cortical components from noisy ones.

$$R_{o/i} = \frac{\text{number of voxels outside the brain}}{\text{number of voxels inside the brain}}. \quad (1)$$

### 2.3.3. Functional Lateralization Index

According to the assumption of relative symmetry in common resting-state brain activities compared to the lateralization of epileptic activities to one hemisphere (Fox et al., 2005), epileptic components are considered to have less relative symmetry about the anterior commissure – posterior commissure (ACPC) plane. This is under the assumption of the right number for extracted components in order to prevent a split in resting-state symmetric ones. After flattening the mirroring voxels of each component about the ACPC plane into two one-dimensional arrays, Pearson's correlation coefficient was used to define the symmetricity between mirroring voxels. The functional lateralization index was defined for each component as follows:

$$LI = 1 - \left| \frac{\sum_{i=1}^{n}(x_i^L - \overline{x^L})(x_i^R - \overline{x^R})}{\sqrt{\sum_{i=1}^{n}(x_i^L - \overline{x^L})^2 \sum_{i=1}^{n}(x_i^R - \overline{x^R})^2}} \right|. \tag{2}$$

*2.3.4. Lateralization Strength*

Following the functional lateralization index, lateralization strength is defined as a measure on significant binary clusters of each component. After thresholding the component by retaining clusters with Z-score > 3.1 and more than ten contiguous voxels, lateralization strength was defined on not-mirrored binary voxels about the ACPC plane as follows:

$$LS = \frac{|\sum X_L - \sum X_R|}{X_L + X_R}. \tag{3}$$

*2.3.5. Local Clustering Coefficient*

Morphological operated clusters were made after opening and closing the sIC clusters with a disk structure of 2 voxels. Main connected componenet was chosen among the connected clusters the one that cosists of the maximum Z-score voxel as the sIC center. The local connectivity matrix was calculated from the fMRI data inside the main cluster of each sIC for later topological measurments. Local clustering coefficient provides a measure of the level of cliquishness or local interconnectedness of a network (Lynall et al., 2010). This measure is defined for the local connectivity of each sIC as the ratio of the sum of existing to the sum of possible connections in the subnetwork:

$$CC = \frac{\sum_{i \neq j}(r_{ij})}{N(N-1)}. \tag{4}$$

### 2.3.6. Local Connectivity Diversity

Local connectivity diversity provides a measure of the heterogeneity of the local network connectivity in each IC. This measure is defined as the unbiased sample variance of all pairwise correlations inside the sIC subnetwork:

$$CD = \frac{1}{N-1} \sum_{j \neq i} (r_{ij} - \bar{r})^2 . \qquad (5)$$

### 2.3.7. Central Network Strength

Central network strength provides a measure of the average level of connectivity between the sIC center and the rest of the sIC voxels.

$$NS = \frac{1}{N-1} \sum_{i \neq c} r_{ic} \quad \text{where} \quad c: sIC\ center . \qquad (6)$$

### 2.3.8. Central Energy

Epileptic temporal activity is expected to have a higher energy caused by interictal hyperactivity. Therefore, the central energy calculates the energy of the fMRI data at the sIC center as a feature for candidate sorting.

$$CE = \sum_t x_c(t)^2 \quad \text{where} \quad c: sIC\ center . \qquad (7)$$

### 2.3.9. Maximum tIC Non-Gaussianity

Another measure for candidate sorting is the maximum non-Gaussianity among the temporal sources of activity in each sIC subregion as that of epileptic discharge signal is expected to be relatively large (Chen et al., 2006). The temporal series of independent sources in the activation region of each candidate component were separated by tICA and the temporal signal with the biggest absolute value of Gaussian deviation (kurtosis) was considered the representative of the epileptogenic temporal activity. The kurtosis of $y$ is defined as below (Hyvarinen and Oja, 2000):

$$Kurtosis\ (y) = E(y^4) - 3(E(y^2)^2) , \qquad (8)$$

where $E(y)$ is the expected value of $y$. After identifying the epileptogenic temporal signal from each candidate component, the eventual epileptic foci can also be localized using seed-based functional connectivity analysis. The above-mentioned feature sets for the 30 sICs of the ten subjects are provided in the Appendix. These features are to be classified into focal and non-focal for epileptic foci localization.

## 2.4. Component Classification

The feature set was classified in a meaningful analytic thresholding procedure with a cut-off of 1 standard deviation (SD) behind the mean of the whole set of sICs and subjects toward the expected direction of the feature. Specifically, features were evaluated based on their maximum power frequency (between 0.01 and 0.1 Hz), voxel-based outside-to-inside ratio (less than the threshold), functional lateralization index (greater than the threshold), lateralization strength (greater than the threshold), local clustering coefficient (less than the threshold), local connectivity diversity (greater than the threshold), central network strength (less than the threshold), central energy (maximum among candidates), and maximum tIC non-Gaussianity (maximum among candidates). The last two features prioritized among the candidates that passed the initial thresholding procedure in each patient. Therefore, one sIC was chosen through this procedure for each patient as the epileptic candidate.

## 3. Results

A total of thirteen patients were recruited in this study. All patients had been diagnosed as surgery candidates with temporal lobe epilepsy and underwent the resection surgery with seizure-free outcome after at least 12-month follow-up. Ten subjects who had post-surgical MRI scans were included for the concordance assessment with the localization results. After preprocessing, sICA was applied to each data and thirty spatially independent components with variance-normalized timecourses were extracted. The feature set was obtained for all 300 components from all patients to calculate the sample means and SDs and hence define the threshold of classification for each feature. The epileptic sIC was extracted for each subject based on the protocol described before. Table 3 shows the feature set of the epileptic sIC in all the subjects. The histograms of each feature and the integral boxplot of the feature set with a normalized range among all 300 components are shown in

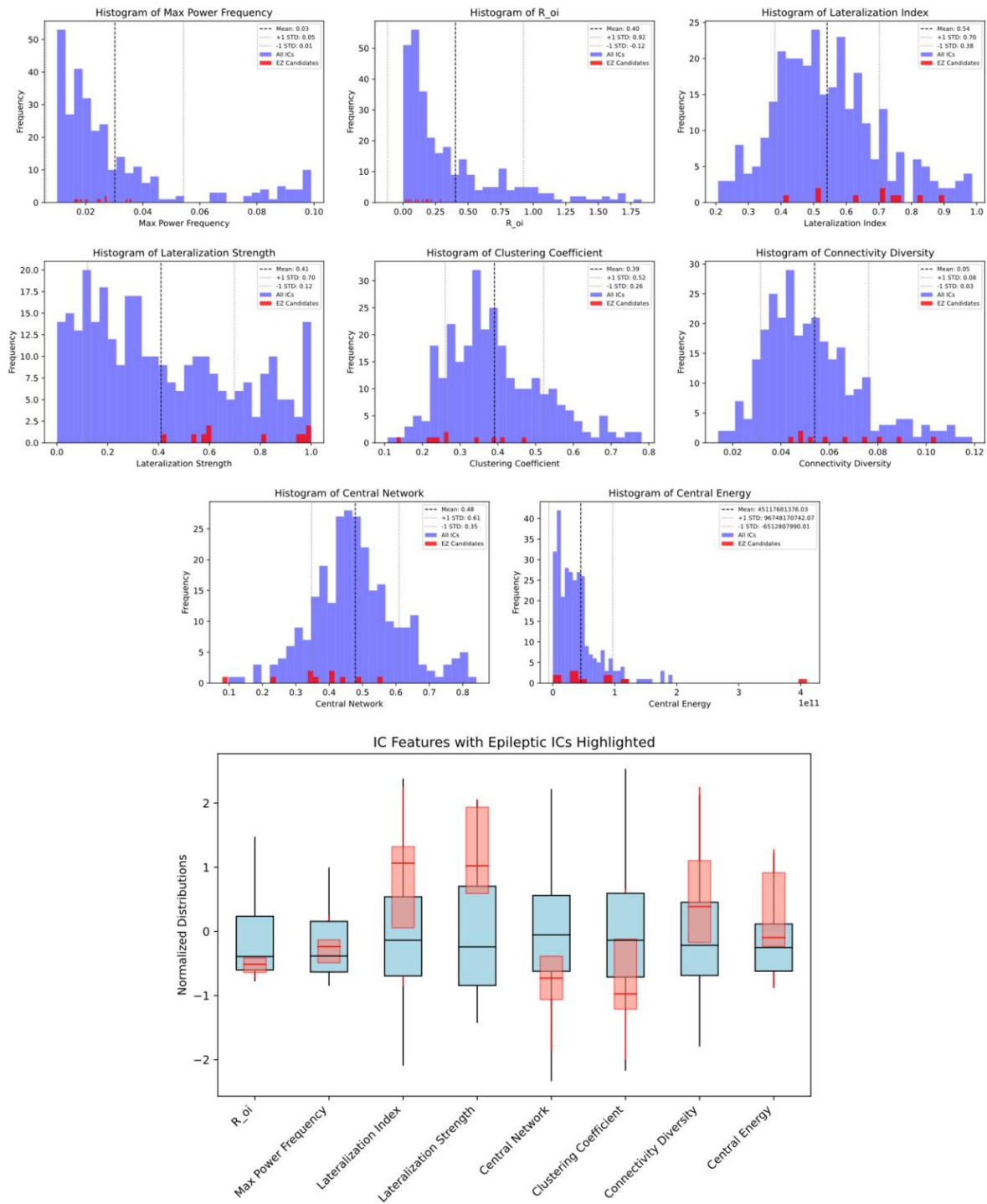

**Figure 2** with those of epileptic ICs being highlighted on each.

The selected components as epileptic tend to be inside the brain, highly lateralized, low in their central network and local clustering coefficient, and high in their local connectivity diversity, central energy, and functional non-Gaussianity. After component classification, the epileptic sIC was clustered to its connected components with Z-score > 3.1 and more than ten contiguous voxels, smoothed by morphological opening and closing with a disk structure of 2 voxels, and reduced to its main connected component that consisted of the maximum Z-score voxel as the resulting SOZ. The SOZs were overlayed on the corresponding post-surgical MRI scan and divided into three levels of concordance. The fully concordant results were spatially aligned with the surgical resection, the partially concordant results were on the same lobe of the resection area without overlap, and the discordant results had their cluster outside the resection lobe. The presented method showed six fully concordant results with precise localization, three partially concordant results with correct lateralization and lobe yet no overlap with the exact resection area, and one discordant result outside the resection lobe.

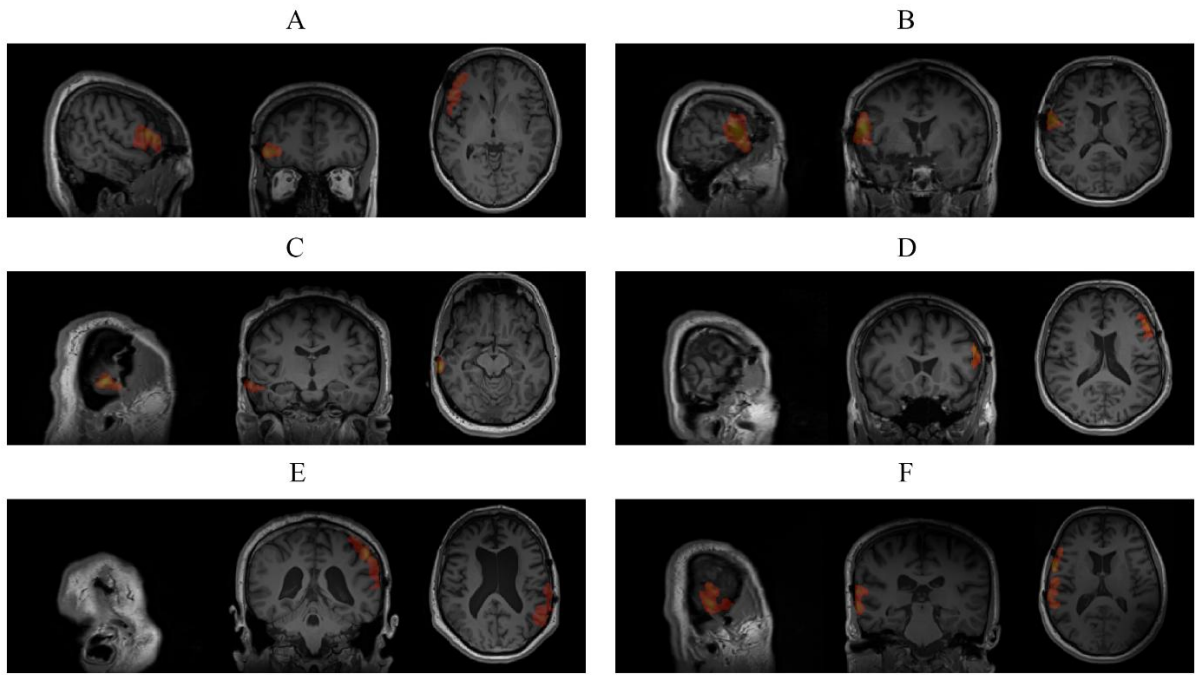

**Figure 3** shows the post-surgical images of six patients with fully concordant results overlayed.

## 4. Discussion

Localization of epileptic foci has become an important step of presurgical evaluation providing a primary perspective of the region of interest in surgery. Although ictal studies and invasive approaches such as iEEG recording have been renowned as the gold standard for this aim, the required time and expense for such methods to investigate the brain and make a final decision might be a problematic issue. Meanwhile, interictal studies and non-invasive approaches may be able to tackle this problem if they enable a quick and extensive investigation of the brain for interictal discharges and provide a reliable protocol to localize the seizure sources throughout the brain. Multi-modal non-invasive data recording, especially simultaneous EEG-fMRI has attracted a great deal of attention over the past years. The idea of combining the temporal resolution of EEG with the spatial resolution of fMRI sounds great to precisely localize a source of abnormal activity in the brain.

However, not all studies in the literature support this idea. Based on the fact that EEG and fMRI measure different phenomena related to epilepsy possibly occurring at different time scales, the relationships between BOLD signal and LFP have been found not always linear but changeable depending on various factors. It has become evident that functional connectivities measured by BOLD and EEG signals have relatively weak correlations, and EEG spikes can either lead to increased, decreased, or unchanged BOLD signals. Moreover, EEG-fMRI methods generally require frequent spikes and a proper expression of epileptic activity on the surface EEG which can be considered a limitation, especially when the epileptic source is in deep brain structures.

The analysis of interictal scalp EEG alone based on ICA and ESI will lead to poor localization results. Direct analysis of interictal rs-fMRI data, on the other hand, has been shown to provide valid information about epileptogenic sources. There are useful tools to analyze and study abnormal activities with specific sources in brain volume from rs-fMRI data using a set of selection criteria on the components obtained from sICA. Temporal analysis is another useful tool to find and confirm a subregion of seizure generation among epileptogenic areas. Moreover, connectivity analysis is a great way to look for the functionally integrated relationships throughout a cluster and among spatially separated brain regions, which can be helpful in finding the areas significantly connected to the seizure onset zone (SOZ) and propagation pathways of epileptic activity. The implementation and comparison of the actual performance of the methods presented on a group of patients has always been valuable research that can lead to the the optimal procedure for localizing epileptic foci during presurgical planning, as well as the development and proposal of new methods in this field.

## 5. Conclusion

In this study, a novel method has been presented to localize epileptic foci based on the analysis of rs-fMRI data alone with no requirement of simultaneously recorded EEG signals. The required functional data for this method is more accessible, cost-efficient, and easier to record compared to other methods based on EEG-fMRI, ictal Video-EEG, MEG, and iEEG. There are fewer assumptions in using rs-fMRI data about the biological features of epilepsy. Moreover, localizing based on rs-fMRI data alone makes the method independent from the depth of epileptic foci and alleviates concerns about the occurrence and detection of epileptic spikes. This method aimed to reach a high spatial accuracy in localizing epileptic foci from interictal data while retaining the reliability of results for clinical usage. After evaluation on a group of patients, the presented method showed promising results compared to the literature highlighting valuable features as SOZ functional biomarkers.

**Acknowledgments**

The authors thank all the volunteers for their contribution. The authors would like to show their gratitude to Dr. Seyed Sohrab Hashemi-Fesharaki (Pars Advanced and Minimally Invasive Medical Manners Research Center, Pars Hospital, Iran University of Medical Sciences, Tehran, Iran) and Dr. Melika Akbarimehr (Neuroscience Research Center, Faculty of Medicine, Qom University of Medical Sciences, Qom, Iran) for sharing their pearls of wisdom during the course of this research. They are also immensely grateful to Dr. Mostafa Asgarinejad (Institute for Cognitive Sciences Studies, Tehran, Iran) for his valuable comments, although any errors are of our own and should not tarnish the reputation of these esteemed individuals.

**Ethics Statement**

This study was carried out in accordance with the recommendations and approvals of the Research Ethics Board (Institutional Review Board, IRB) of the Tehran University of Medical Sciences with written informed consent from all subjects in accordance with the Declaration of the Cognitive Sciences & Technologies Council. We also confirm that we have read the Journal's position on issues involved in ethical publication and affirm that this report is consistent with those guidelines.

**Author Contributions**

SMS, EE, AF, and HSZ conceived of the presented idea. SMS and EE developed the theory and performed the computations. Material preparation, data collection, and analysis were performed by SMS, EE, AF. The first draft of the manuscript was written by SMS and all authors commented on the manuscript. JM and HSZ verified the analytical methods. The visualization and validation were done by SMS. All authors provided critical feedback and helped shape the research, analysis, and manuscript. All authors read and approved the final manuscript. HSZ supervised the project.

**Funding**

This research did not receive any specific grant from funding agencies in the public, commercial, or not-for-profit sectors.

**Conflict of Interest**

The authors declare no conflicts of interest related to this work.

**Competing Interests**

The authors declare that they have no known competing financial interests or personal relationships that could have appeared to influence the work reported in this paper.

**Data Availability Statement**

The dataset used in this study will be made available upon request to the authors.

**Code Availibility Statement**

The code implementing the entire process of the presented method is available at https://github.com/smsadjadi/Spatio-Temporal-Component-Classification-for-Localizing-Seizure-Onset-Zone.

Table 1. Studies presenting epileptic foci localization methods using EEG, fMRI, and EEG-fMRI data and including seizure-free surgical outcome data for evaluation.

| Study | Year | Sample Size | Method Data | Method | Accuracy / Results |
|---|---|---|---|---|---|
| (Brodbeck et al., 2010) | 2010 | 10 | EEG | Eelctrical source imaging (ESI) using LAURA on the IEDs. | 8/9 (88%) within the resection margins. |
| (Thornton et al., 2010) | 2010 | 34 | EEG-fMRI | Conventional EEG-fMRI method using the extracted timing of IEDs convolved with canonical hemodynamic response function (HRF) as the regressor for GLM on fMRI data. | 10/34 (30%) had surgical resection and significant activation on EEG-fMRI, 7/10 were seizure-free following surgery, and 6/7 had concordant results with resection. |
| (Grouiller et al., 2011) | 2011 | 23 | EEG-fMRI | The EEG voltage map of the IED template was correlated with Intra-MRI EEG voltage maps to construct the epileptic activity for further GLM analysis. | 10/18 (55%) fully concordant and 4/18 (22%) were moderately concordant to the postoperative areas with seizure freedom. |
| (Jann et al., 2012) | 2012 | 10 | EEG-fMRI | Three different threshold criteria were applied to detect hemodynamic responses to the IEDs: peak activation (criterion 1), fixed threshold at P<.05 corrected for multiple comparison (criterion 2), and fixed numbers of activated voxels (4000±200) within the brain (criterion 3). | 5/10 (50%) concordance with criterion 1, 6/10 (60%) concordance with criterion 2, 8/10 (80%) concordance with criterion 3. |
| (Pouliot et al., 2012) | 2012 | 3 | EEG-fMRI | Non-linear hemodynamic responses using the second-order expansion of the Volterra kernel with epileptic spikes as time-dependent inputs and BOLD, oxyhemoglobin (HbO), and deoxyhemoglobin (HbR) time series at a certain fMRI voxel as the outputs. | 3/3 (100%) concordance of significant non-linearities with the epileptic foci and negative BOLD response regions. |
| (An et al., 2013) | 2013 | 35 | EEG-fMRI | Conventional method with combined t map of four HRFs peaking at three, five, seven, and nine seconds. | 10/35 (29%) fully lobe concordant, 9/35 (26%) partially lobe concordant, 5/35 (14%) partially lobe discordant, 11/35 (31%) fully lobe discordant. |
| (van Houdt et al., 2015) | 2015 | 8 | fMRI | sICA was applied to two fMRI data epochs with and without visible IEDs separately and the epileptic sIC was found using spatial correlation with the resection area and EEG-fMRI correlation patterns. | 7/8 (88%) remarkable resemblance between the epileptic sICs in the two states suggesting that epilepsy-related sICs are not dependent on the presence of IEDs. |
| (Hunyadi et al., 2015) | 2015 | 12 | EEG-fMRI | Most correlated sICs with resection areas were obtained from EEG tICA and fMRI sICA, and correlation coefficients were calculated for all possible pairs of EEG-eICs convolved with HRF and fMRI-eICs. | 3/12 (25%) epileptic sICs were matched between EEG and fMRI and overlapped to the epileptic zone. |
| (Zhang et al., 2015) | 2015 | 9 | fMRI | sICA was applied to fMRI data extracting 30 sICs and the epileptic sIC was found using biophysical constraints, temporal features, and lateralization index. | 7/9 (78%) lobe concordant results. |
| (Coan et al., 2016) | 2016 | 30 | EEG-fMRI | The regressor proposed in (Grouiller et al., 2011) plus the conventional regressor were used together in GLM analysis. | 81% sensitivity and 79% specificity of results to identify patients with good surgical outcome. |
| (Centeno et al., 2017) | 2017 | 53 | EEG & EEG-fMRI | EEG-fMRI global maxima (GM) along with ESI. | 17/53 (32%) localized by ESI, 11/53 (21%) localized by GM, and 11/53 (21%) localized by both with the mean distance of 14.6 mm in their maxima. |
| (Maziero et al., 2018) | 2018 | 18 | EEG-fMRI | 2dTCA presented for mapping the seizure onset zone. | 13/18 (72%) concordant results not confined to the presence of IEDs. |
| (Chaudhary et al., 2021) | 2021 | 8 | iEEG-fMRI | 38 different topographic IEDs were classified and extracted from iEEG and BOLD changes associated with individual IED classes were assessed over the whole brain using GLM. | 27/38 (71%) IED classes resulted in concordant BOLD maps. |

Table 2. Clinical and electrophysiological information of the patients.

| No. | Frequency | Onset (years) | Handedness | Semiology (Salient Features) | Ictal EEG (LTM) | Epileptogenic Zone (LTM) | Irritative Zone (LTM) | MRI (MTS) | Laterality | Outcome |
|---|---|---|---|---|---|---|---|---|---|---|
| Sub001 | 1/m | 28 | R | Behavioral arrest with oral automatisms; verbalization (R) | Rhythmic alpha & theta activity R (T>F) | R (T) | R>L (F<T) | R | R | Engel I |
| Sub002 | - | 13 | R | Bilateral limb automatisms | Rhythmic theta activity L=R (T) | L=R (F<T) | (R>L) (T) | R | R | Engel I |
| Sub003 | 7-12/w | 13 | R | Experiential aura; behavioral arrest; right limb dystonia (L) | Rhythmic theta activity L (T) | L (T) | R > L (T) | L | L | Engel I |
| Sub004 | 4/m | 0.6 | R | Behavioral arrest with staring; right limb dystonia; verbalization | Rhythmic delta activity R (T>F), L (T) Rhythmic theta activity R (T) Rhythmic delta activity L>R (T) | R (T) | R (T) | R | R | Engel I |
| Sub005 | 0.3-1/m | 22 | R | Left versive motion; left limb dystonia (R) | Rhythmic alpha activity R>L (T) | R (T) | R (T) | R | R | Engel I |
| Sub006 | 4/w | 3 | R | Staring with oral automatisms; right versive motion, right facial clonic activity (L) | Rhythmic theta activity L>R (T) | L (T) | - | L | L | Engel I |
| Sub007 | 1-4/w | 19 | R | Behavioral arrest; left limb automatism, right versive motion (L) | Rhythmic theta activity L>R (T) | L (T) | L (T) | L | L | Engel I |
| Sub008 | - | - | - | Behavioral arrest with staring | - | R (T) | - | - | R | Engel I |
| Sub009 | 7-12/m | 4 | L | Experiential aura; behavioral arrest; Right Arm Dystonia | Rhythmic theta activities L>R (T) | L (T) | L (T) | L | L | Engel I |
| Sub010 | 2-3/w | 2 | R | Behavioral arrest with staring and oral automatism, spitting; left limb dystonia | Rhythmic theta R>L (T) Rhythmic delta R>L (T>F) | R (T) | (R>L) (T) | R>L | R | Engel I |
| Sub011 | 2-3/m | 14 | R | Behavioral arrest with staring | - | L (T) | - | L | L | Engel I |
| Sub012 | 1-4/m | 11 | R | Behavioral arrest with blinking and oral automatisms | Rhythmic theta activities L >R (T) | - | L | L | L | Engel I |
| Sub013 | 1-15/m | - | R | Behavioral arrest with blinking; left limb dystonia; ictal laughter (R) | Rhythmic alpha activity R>L (T) | R (T) | R (T) | R | R | Engel I |

LTM: Long-Term Monitoring; MTS: Mesial Temporal Sclerosis.

Table 3. The feature set of the epileptic sICs.

| Subject ID | sIC Number | sIC Center | $R\_o/i$ | Max Power Frequency | Lateralization Index | Lateralization Strength | Lateralization Side | Central Network | Clustering Coefficient | Connectivity Diversity | Central Energy | Max tIC Non-Gaussianity |
|---|---|---|---|---|---|---|---|---|---|---|---|---|
| 1 | 16 | (60.0, 22.0, 12.0) | 0.04 | 0.036 | 0.756 | 0.994 | Right | 0.482 | 0.387 | 0.043 | 95325948678.60340 | 315.003133545483 |
| 2 | 17 | (62.0, -8.0, 22.0) | 0.018 | 0.025 | 0.51 | 0.411 | Right | 0.56 | 0.475 | 0.049 | 32458319850.63030 | 315.003133729914 |
| 3 | 13 | (-64.0, -48.0, 26.0) | 0.092 | 0.017 | 0.637 | 0.588 | Left | 0.348 | 0.231 | 0.053 | 37430046193.99810 | 315.003134102643 |
| 4 | 15 | (38.0, -52.0, 64.0) | 0.061 | 0.02 | 0.718 | 1 | Right | 0.237 | 0.213 | 0.104 | 110812658071.97300 | 315.003133634571 |
| 5 | 27 | (58.0, -56.0, 32.0) | 0.156 | 0.018 | 0.705 | 0.543 | Right | 0.433 | 0.339 | 0.074 | 34834436821.58160 | 315.003132862966 |
| 6 | 17 | (-54.0, 26.0, -6.0) | 0.189 | 0.027 | 0.829 | 0.944 | Left | 0.363 | 0.266 | 0.049 | 42803728111.47780 | 315.003133696123 |
| 7 | 26 | (-50.0, -44.0, 56.0) | 0.184 | 0.016 | 0.9 | 0.976 | Left | 0.336 | 0.238 | 0.08 | 409992669118.26000 | 315.003130300942 |
| 8 | 4 | (58.0, 16.0, -2.0) | 0.217 | 0.034 | 0.521 | 0.578 | Right | 0.402 | 0.261 | 0.059 | 83131683704.61460 | 315.003130164952 |
| 9 | 15 | (-22.0, -44.0, 82.0) | 0.295 | 0.024 | 0.742 | 0.604 | Left | 0.08 | 0.131 | 0.089 | 12609534.58141 | 315.003126433931 |
| 10 | 10 | (64.0, 20.0, 8.0) | 0.115 | 0.027 | 0.407 | 0.805 | Right | 0.41 | 0.417 | 0.066 | 8724295223.41954 | 315.003133076034 |

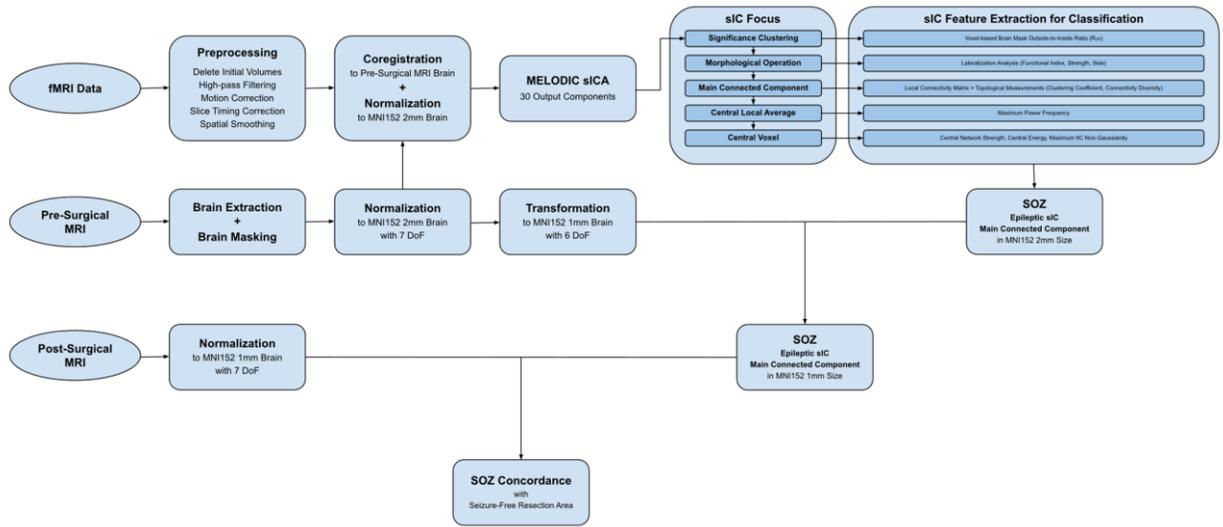

Figure 1. Diagram of the presented method algorithm.

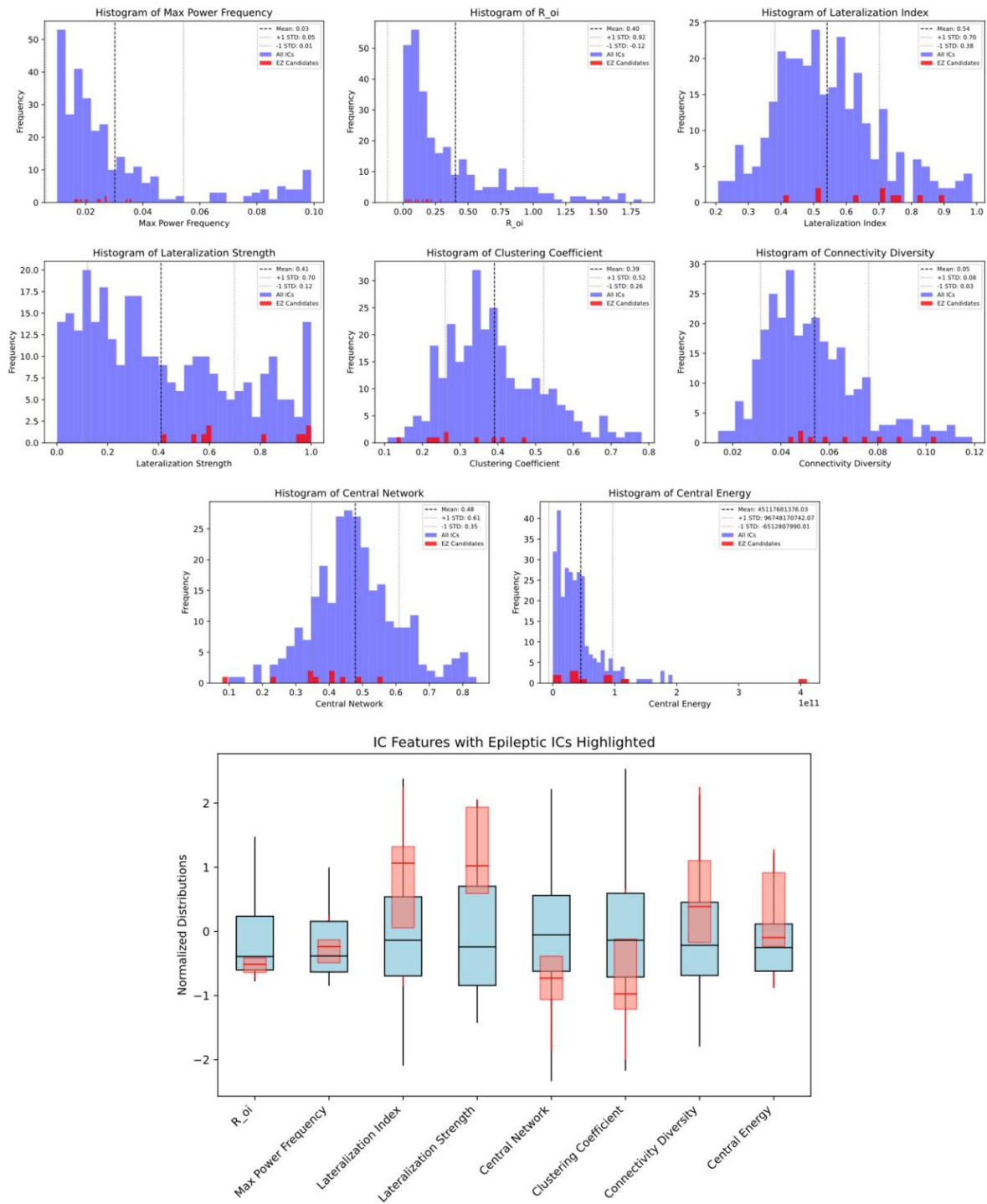

**Figure 2.** Histogram of features among all sICs and subjects with the epileptic sICs being highlighted (Top) and the boxplot of the feature set with a normalized range among all sICs and subjects with the epileptic sICs being highlighted (Bottom).

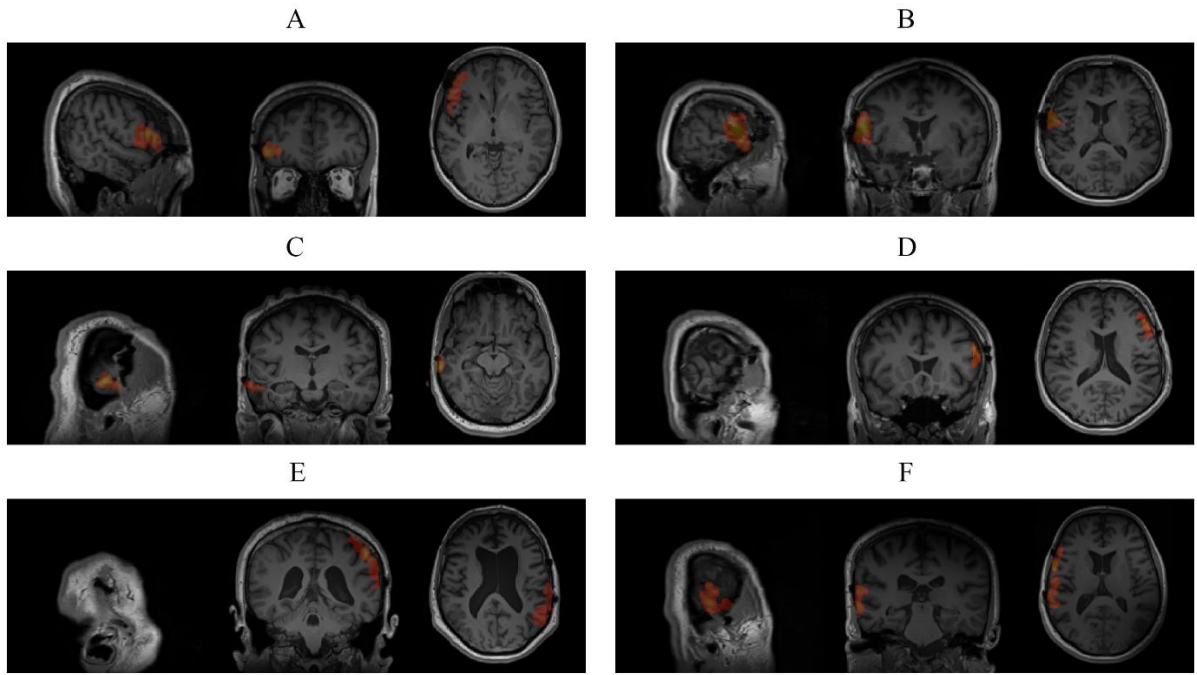

**Figure 3.** The post-surgical MRI with overlaid localized SOZ cluster on resection area using the presented method: [A] subject 1 with behavioral arrest with oral automatisms and verbalization showing rhythmic alpha and theta activity at right temporal lobe (T4>F8) in ictal EEG; [B] subject 2 with bilateral limb automatisms showing rhythmic theta activity at both temporal lobes (T>F) but right irritative zone in ictal EEG; [C] subject 5 with left versive motion and left limb dystonia showing rhythmic alpha activity at right temporal lobe (T4>T3) in ictal EEG; [D] subject 6 with staring with oral automatisms, right versive motion, and right facial clonic activity showing rhythmic theta activity at left temporal lobe (T3>T4) in ictal EEG; [E] subject 7 with behavioral arrest, left limb automatism, and right versive motion showing rhythmic theta activity at left temporal lobe (T3>T4) in ictal EEG; [F] subject 8 with behavioral arrest with staring.